\documentclass[a4paper]{jpconf}
\usepackage{amsfonts,amsmath,amssymb,amsthm,bbm}
\usepackage{graphicx,hyperref}

\newcommand{\N}{{\mathbb N}}

\newcommand{\cE}{{\mathcal E}}

\newcommand{\cH}{{\mathcal H}}

\newcommand{\SU}{\mathrm{SU}}

\newcommand{\U}{\mathrm{U}}

\def\inv{{\textrm{Inv}}}

\def\cHNJ{\cH_N^{(J)}}
\def\cHN{\cH_N}

\newcommand{\be}{\begin{equation}}
\newcommand{\ee}{\end{equation}}
\newcommand{\beq}{\begin{eqnarray}}
\newcommand{\eeq}{\end{eqnarray}}
\newcommand{\bea}{\begin{eqnarray}}
\newcommand{\eea}{\end{eqnarray}}
\newcommand{\nn}{\nonumber}

\newcommand{\su}{{\mathfrak su}}
\renewcommand{\sl}{{\mathfrak sl}}

\renewcommand{\u}{{\mathfrak u}}

\newcommand{\ra}{\rangle}

\def\nn{\nonumber}

\def\Ea{E^{(\alpha)}}
\def\Eb{E^{(\beta)}}
\def\Fa{F^{(\alpha)}}
\def\Fb{F^{(\beta)}}
\def\dag{^\dagger}

\begin{document}
\title{$\U(N)$ invariant dynamics for a simplified loop quantum gravity model}

\author{Enrique F. Borja}
\address{Institute for Theoretical Physics III, University of
Erlangen-N\"{u}rnberg, Staudtstra{\ss}e 7, D-91058 Erlangen (Germany).}
\address{Departamento de F\'{\i}sica Te\'{o}rica and IFIC, Centro Mixto Universidad de
Valencia-CSIC. Facultad de F\'{\i}sica, Universidad de Valencia,
Burjassot-46100, Valencia (Spain).}
\ead{efborja@theorie3.physik.uni-erlangen.de}

\author{Jacobo D\'{\i}az-Polo}
\address{Institute for Gravitation and the Cosmos \& Physics
Department, Penn State, University Park, PA 16802-6300, U.S.A.}
\ead{jacobo@gravity.psu.edu}

\author{I\~{n}aki Garay}
\address{Institute for Theoretical Physics III, University of
Erlangen-N\"{u}rnberg, Staudtstra{\ss}e 7, D-91058 Erlangen (Germany).}
\ead{igael@theorie3.physik.uni-erlangen.de}

\author{Etera R. Livine}
\address{Laboratoire de Physique, ENS Lyon, CNRS-UMR 5672, 46
All\'ee d'Italie, Lyon 69007, France.}
\ead{etera.livine@ens-lyon.fr}

\begin{abstract}
The implementation of the dynamics in Loop Quantum Gravity (LQG) is
still an open problem. Here, we discuss a tentative dynamics for the
simplest class of graphs in LQG: Two vertices linked with an
arbitrary number of edges. We use the recently introduced $\U(N)$
framework in order to construct $\SU(2)$ invariant operators and
define a global $\U(N)$ symmetry that will select the
homogeneous/isotropic states. Finally, we propose a Hamiltonian
operator invariant under area-preserving deformations of the
boundary surface and we identify possible connections of this model
with Loop Quantum Cosmology.
\end{abstract}

\section{Introduction}

Loop quantum gravity (LQG) is now a well-established approach to
quantum gravity \cite{librothomas}. It provides a non-perturbative
mathematical formulation of the kinematical sector of the theory.
The Hilbert space is generated by spin-networks: wave functions
defined over oriented graphs whose edges are labeled by irreducible
representations of the $\SU(2)$ group, and with intertwiners
($\SU(2)$ invariant tensors) on its vertices. Despite the several
advances that have taken place in this field, one of the main
challenges faced by the theory is the implementation of the
dynamics. Our goal is to focus on a specific model in order to
propose a suitable Hamiltonian for it.

Rovelli and Vidotto introduced a simple model based on a graph with
2 vertices and 4 edges \cite{carlo1}. They found that this model
leads to a physical framework very similar to Loop Quantum Cosmology
(LQC). We generalize this model and implement the LQG dynamics on a
graph with 2 vertices joined by an arbitrary number $N$ of edges
\cite{2vertex}.

We use the recently developed $\U(N)$ framework for
$\SU(2)$-intertwiners \cite{un1,un2,un3} in order to study the
Hilbert space of spin  networks on the 2-vertex graph. The operators
of the $\U(N)$ formalism act on our Hilbert space and we identify a
global $\U(N)$ symmetry generated by operators acting on the coupled
system of the two vertices which reduces the full space of arbitrary
spin network states to a space of homogeneous/isotropic states.
Finally, we introduce global $\U(N)$-invariant operators and use
them to propose a $\U(N)$-invariant Hamiltonian operator.

\section{The $\U(N)$ framework}

This framework was introduced in a series of papers
\cite{un1,un2,un3}. Intertwiners with $N$ legs are
$\SU(2)$-invariant states in the tensor product of $N$ (irreducible)
representations of $\SU(2)$. Then the  basic tool used is the
Schwinger representation of the $\su(2)$ Lie algebra in terms of a
pair of harmonic oscillators. Since we would like to describe the
tensor product of $N$ $\SU(2)$-representations, we will need $N$
copies of the $\su(2)$-algebra  and thus we consider $N$ pairs of
harmonic oscillators $a_i,b_i$ with $i$ running from 1 to $N$.

We can identify $\SU(2)$ invariant operators acting on pairs of
(possibly equal) legs $i,j$ \cite{un1,un3}:
\be
E_{ij}=a\dag_ia_j+b\dag_ib_j, \quad (E_{ij}\dag=E_{ji}),\quad\qquad
F_{ij}=a_i b_j - a_j b_i,\quad (F_{ji}=-F_{ij}).\nn
\ee
The operators $E$ form a $\u(N)$-algebra
\be
[E_{ij},E_{kl}]= \delta_{jk}E_{il}-\delta_{il}E_{kj},\nn
\ee
and with the operators $F,F\dag$ form a closed algebra. This is why
this formalism has been dubbed the $\U(N)$ framework for LQG. Notice
that the diagonal operators give the energy on each leg,
$E_{ii}=E_i$. Then the value of the  total energy $E\,\equiv \sum_i
E_i$ gives twice the sum of all spins $2\times\sum_i j_i$, i.e twice
the total area around the vertex in the context of LQG.

The $E_{ij}$-operators change the energy/area carried by each leg,
while still conserving the total energy, while the operators
$F_{ij}$ (resp. $F\dag_{ij}$) will decrease (resp. increase) the
total area $E$ by 2:
\be
[E,E_{ij}]=0,\qquad [E,F_{ij}]=-2F_{ij},\quad
[E,F\dag_{ij}]=+2F\dag_{ij}.\nn
\ee
This suggests to decompose the Hilbert space of $N$-valent
intertwiners into subspaces of constant area:
\be
\cH_N=\bigoplus_{\{j_i\}} \inv\left[\otimes_{i=1}^NV^{j_i}\right]
=\bigoplus_{J\in\N}\bigoplus_{\sum_ij_i=J}
\inv\left[\otimes_{i=1}^NV^{j_i}\right] =\bigoplus_J \cH_N^{(J)},\nn
\ee
where $V^{j_i}$ denotes the Hilbert space of the irreducible
$\SU(2)$-representation of spin $j_i$, spanned by the states  of the
oscillators $a_i,b_i$ with fixed total energy $E_i=2j_i$. Besides,
it was proven \cite{un2} that each subspace $\cHNJ$ of $N$-valent
intertwiners with fixed total area $J$ carries an irreducible
representation of $\U(N)$ generated by the $E_{ij}$ operators.

Then the operators $E_{ij}$ allow to navigate from state to state
within each subspace $\cHNJ$. On the other hand, the operators
$F_{ij},\,F\dag_{ij}$ allow to go from one subspace $\cHNJ$ to the
next $\cHN^{(J\pm 1)}$, thus endowing the full space of $N$-valent
intertwiners with a Fock space structure with creation operators
$F\dag_{ij}$ and annihilation operators $F_{ij}$. It is worth to
point out that the operators $E_{ij},F_{ij},F\dag_{ij}$ satisfy
certain quadratic constraints \cite{2vertex} that look a lot like
constraints on the multiplication of two matrices, one of them
hermitian and the other antisymmetric \cite{return}.

\section{The 2 vertex model}

\begin{figure}[h]
\begin{center}
\includegraphics[height=38mm]{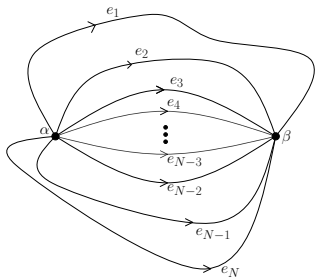}
\caption{The 2-vertex graph with vertices $\alpha$ and $\beta$ and
the $N$ edges linking them.} \label{2vertexfig}
\end{center}
\end{figure}

We consider the simplest class of non-trivial graphs for spin
network states in Loop Quantum Gravity: a graph with two vertices
linked by $N$ edges, as shown in fig.\ref{2vertexfig}. This is a
generalization of the simplest model (2 vertices and 4 links)
introduced by Rovelli and Vidotto in \cite{carlo1} and which was
shown to be related to models of quantum cosmology \cite{carlo2}.

There are matching conditions \cite{un2} imposing that each edge
carries a unique $\SU(2)$ representation, thus the spin on that edge
must be the same as seen from $\alpha$ than from $\beta$ i.e.
$j_i^\alpha=j_i^\beta$. This translates into the fact that the
oscillator energy for $\alpha$ on the leg $i$ must be equal to the
energy for $\beta$ on its $i$-th leg:
\be
\cE_i\,\equiv\,\Ea_i -\Eb_i \,=\,0.\nn
\ee
We can see that the matching conditions $\cE_k$ generate a $\U(1)^N$
symmetry, so they are part of a larger $\U(N)$ symmetry algebra.
Indeed, it is possible to introduce the more general operators:
\be
\cE_{ij}\,\equiv\, \Ea_{ij}-\Eb_{ji}
\,=\,\Ea_{ij}-(\Eb_{ij})\dag.\nn
\ee
that form a $\U(N)$ algebra and that reduce to the matching
conditions in the case $i=j$.

Now, we can show \cite{2vertex} that by imposing the global
$\U(N)$-invariance just defined on our 2-vertex system, we obtain a
single state $|J\ra$ for every total boundary area $J$. Thus, the
$\U(N)$ invariance is restricting our system to states which are not
sensitive to area-preserving deformations of the boundary between
$\alpha$ and $\beta$. They are isotropic in the sense that all
directions (i.e. all edges) are equivalent and the state only
depends on the total boundary area, and they are homogeneous in the
sense that the quantum state is the same at every point of space,
i.e. at $\alpha$ and $\beta$.

In the following, we will restrict our study to this global $\U(N)$
invariant Hilbert space and we will propose a dynamics for this
system. Studying the structure of the $\U(N)$ invariant operators
acting on this space, we propose the simplest and most natural
ansatz for a Hamiltonian operator:
\be
H \,\equiv\, \lambda\sum_{ij}\Ea_{ij}\Eb_{ij}+ \left(\sigma
\sum_{ij}\Fa_{ij}\Fb_{ij} +\bar{\sigma}
\sum_{ij}F^{\alpha\dagger}_{ij}F^{\beta\dagger}_{ij}\right)\,,\nn
\ee
where the coupling $\lambda$ is real while $\sigma$ can be complex a
priori, so that the operator $H$ is Hermitian. In fact, this is the
most general $\U(N)$ invariant Hamiltonian (allowing only elementary
changes in the total area), up to a renormalization by a
$E$-dependent factor.

The action of this Hamiltonian over a state $|J\ra$ is known, and we
can also study the spectral properties of this operator. It turns
out that it shares several mathematical analogies with the evolution
operator used in Loop Quantum Cosmology
\cite{lqc_spectrum3,lqc_spectrum2}. From this perspective, the
2-vertex model has a natural cosmological interpretation and
exhibits a ``big bounce'' behavior avoiding the big bang singularity
as in LQC.

\section{Conclusions}

The $\U(N)$ framework recently introduced in \cite{un1,un2,un3}
represents a new and refreshing way to tackle several issues in the
Loop Quantum Gravity framework.

In this work we have described a very simple model for LQG, based on
the 2-vertex graph described above.  We have the space of
intertwiners for each vertex, plus some matching conditions coupling
the two intertwiners by imposing that each edge $i$ of the graph
carries a unique spin label $j_i$ as seen from both vertices. These
matching conditions form the Cartan subalgebra of a larger
``global'' $\u(N)$ algebra acting on both vertices, and we have seen
that the invariance under this global $\U(N)$ symmetry implies the
restriction to isotropic and homogeneous states $|J\ra$ at the
quantum level. This provides our $\U(N)$ symmetry with a very
concrete physical interpretation. Indeed it could be viewed as a key
step towards a full understanding between the general Loop Quantum
Gravity (LQG) framework and the symmetry-reduced Loop Quantum
Cosmology (LQC).

We have focused on the global $\U(N)$ invariant space of
spin-network states. Using suitable $\U(N)$ invariant operators we
have proposed a dynamics for this system. The defined $\U(N)$
invariant Hamiltonian allows a direct comparison with the
Hamiltonian constraints used in LQC. This fact reinforces the
possible connection between this kind of models and the Loop Quantum
Cosmology framework.

\section*{Acknowledgments}

This work was in part supported by the Spanish MICINN research
grants FIS2008-01980, ESP2007-66542-C04-01 and FIS2009-11893. JD is
supported by the NSF grant PHY0854743, The George A. and Margaret M.
Downsbrough Endowment and the Eberly research funds of Penn State.
IG is supported by the Department of Education of the Basque
Government under the ``Formaci\'{o}n de Investigadores'' program. EL is
partially supported by the ANR ``Programme Blanc" grants LQG-09 and
by the ESF short visit travel grant 3595.

\end{document}